\pdfoutput=1

\RequirePackage{ifpdf}
\documentclass{JINST}

\usepackage{booktabs}
\usepackage{amsmath} 
\usepackage{subcaption}

\title{Intrinsic Pixel Size Variation in an LSST Prototype Sensor}

\author{M. A. Baumer\thanks{Corresponding author.}~,
A. Roodman\\
Kavli Institute for Particle Astrophysics and Cosmology,\\
SLAC National Accelerator Laboratory, \\
Stanford University Department of Physics \\
2575 Sand Hill Road MS 29 \\
Menlo Park, CA 94025 USA\\

E-mail: \email{mbaumer@stanford.edu}}

\abstract{The ambitious science goals of the Large Synoptic Survey Telescope (LSST) have motivated a search for new and unexpected sources of systematic error in the LSST camera. Flat-field images are a rich source of data on sensor anomalies, although such effects are typically dwarfed by shot noise in a single flat field. After combining many ($\sim 500$) such images into `ultraflats' to reduce the impact of shot noise, we perform photon transfer analysis on a pixel-by-pixel basis and observe no spatial structure in pixel linearity or gain at light levels of 100 ke$^-$ and below. At 125 ke$^-$, a columnar structure is observed in the gain map---we attribute this to a flux-dependent charge transfer inefficiency. We also probe small-scale variations in effective pixel size by analyzing pixel-neighbor correlations in ultraflat images, where we observe clear evidence of intrinsic variation in effective pixel size in an LSST prototype sensor near the $\sim .3\%$ level.}


\begin{document}

\section{Introduction}\label{sec:intro}

With the great promise of LSST science come unprecedented demands on camera performance \cite{science}. These demands have made it necessary to investigate heretofore unconsidered sources of systematic errors within the LSST camera.

The search for new sensor systematics in prototype LSST CCDs has been the subject of much work over the past few years. The so-called `brighter-fatter effect', which induces a flux-dependent point-spread function in astronomical applications, has been successfully modeled in terms of pixel-neighbor correlations \cite{antilogus}. This model has been used by the Dark Energy Survey to derive a first-order correction for this effect in DECam CCDs. Without proper consideration of this sensor effect, the Dark Energy Survey would have been unable to reach its sensitivity goal for Stage III dark energy measurements \cite{gruen}.

Given that sensor effects have been shown to have significant impact on science in modern cosmological surveys, we wish to scrutinize sensor behavior more and more closely, until we can characterize and correct for all known effects (or show that their impact on LSST Stage IV dark energy science can be neglected). One possible new source of error comes from small-scale variations in pixel sensitivity, classically referred to as pixel response non-uniformity (PRNU). The assumption that PRNU is primarily caused by local variation in the quantum efficiency (QE) of CCD pixels is the foundation of the widespread use of flat fielding corrections in modern astronomy. However, more recent work has indicated that intrinsic pixel size variation can also contribute to PRNU \cite{smith}. Small deviations from the nominal dimensions of $10 \times 10$ microns might arise from lateral electric fields produced by impurities within the silicon bulk of a CCD \cite{stubbs}. It is important to note that the distortions induced by such pixel boundary shifts are not properly corrected by naive flat-fielding, because such a correction normalizes the observed flux per unit area, rather than the flux per pixel, which is the desired calibration.

\section{Data and Pre-processing}

The scale of pixel-to-pixel variation in effective area, if present in a sensor, is expected to be small relative to single-exposure shot noise. Therefore, to investigate these effects, we need to co-add a significant number of flat-field images in order to clearly observe a potential effect. In this study, we constructed `ultraflat' images by combining 500 flat-field exposures at each of four different light levels taken with an e2v LSST prototype sensor (full well $= 170 \text{ ke}^-$ \cite{personal}) from the Harvard sensor testing lab. The illumination configurations are described in Table \ref{table:flats}.

\begin{table}[tbp]
\centering
\begin{tabular}{ccc}
\toprule
Light Level (e$^-$) & PRNU (\%) & Residual Shot Noise (\%) \\
\midrule
25,000 & .28 & .028 \\
75,000 & .35 & .016 \\
100,000 & .36 & .014 \\
125,000 & .35 & .013 \\
\bottomrule
\end{tabular}
\caption{The measured PRNU and shot noise contamination ($\sigma_{shot} = \sqrt{\Phi / N}$) expressed as a percentage of mean flux $\Phi$ for ultraflats at four different light levels. The co-addition of 500 exposures brings the shot noise contamination down to a factor of 10-20 below the PRNU at each light level.}
\label{table:flats}
\end{table}

\begin{figure}[tbp] 
\centering
\includegraphics[width=3in]{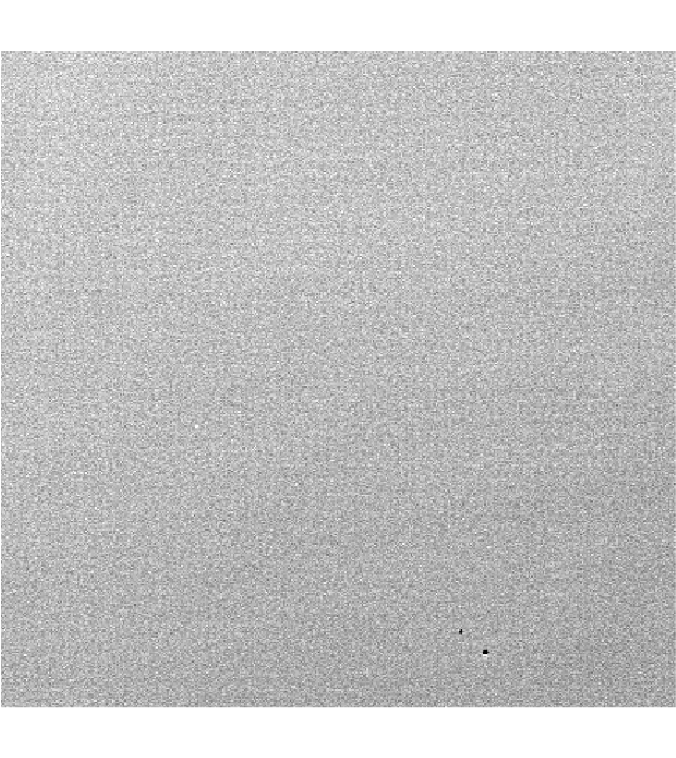}
\caption{A $400 \times 400$ pixel ($\sim20\%$ of a single amplifier segment) cutout from an ultraflat taken at the 75ke$^-$ light level. This sensor exhibits remarkable cosmetic quality: in this representative example, only two small dead regions are visible near the bottom of the image.}
\label{fig:cutout}
\end{figure}

Each image is overscan- and bias-corrected, and the mean flux level of each image is adjusted to be consistent across the 500 frames, to correct for small variations in light level due to shutter timing and lamp instability. Since we are primarily interested in size variation on the few-pixel scale, we apply a high-pass filter to remove large-scale illumination variation by subtracting a smoothed image from the data to highlight the small-scale structure of the flat fields. Finally, to avoid potential sensor edge effects, we crop a 50 pixel border from each amplifier of the co-added ultraflat. A representative cutout from a 75ke$^-$ ultraflat is shown in Figure \ref{fig:cutout}, illustrating the sensor's excellent cosmetics.

\section{Single-pixel Photon Transfer}

\begin{figure}[tbp]
\centering
\begin{subfigure}[b]{0.45\textwidth}
\centering
\includegraphics[height=2.1in]{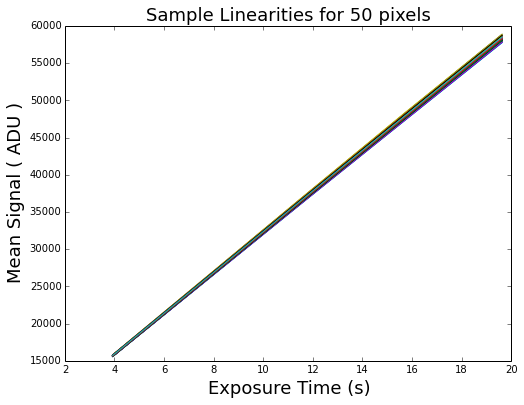}
\caption{Sample linearity plots of mean signal vs. exposure time for 50 random pixels. Each pixel has a linear response that is matched to the others to within .4\%.}
\label{fig:lin}
\end{subfigure}
\hfill
\begin{subfigure}[b]{0.45\textwidth}
\centering
\includegraphics[height=2.1in]{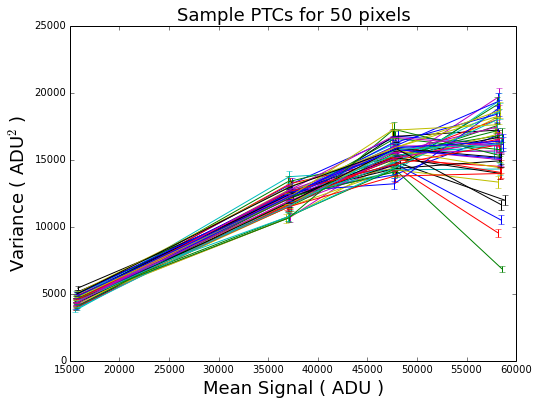}
\caption{Sample PTCs for same 50 pixels plotted in previous panel. Several pixels exhibit a significant and unexpected loss of variance at the 125ke$^-$ light level, despite normal linearity.}
\label{fig:ptc}
\end{subfigure}
\hfill
\caption{Linearity and PTC plots for a representative sample of pixels in our single-pixel PTC analysis.}
\label{fig:single_ex}
\end{figure}

The photon transfer curve (PTC) is a common technique for characterizing the noise performance of CCD sensors. The gain can be measured from the inverse slope of a variance vs. mean signal plot, and sensor linearity can be measured as the slope of a mean flux vs. exposure time plot. Typically this analysis is done by computing the mean and variance across the focal plane at different light levels and making a single plot that characterizes overall sensor performance. With ultraflats, we have sufficient statistics do this analysis on a pixel-by-pixel basis, computing average flux and variance for individual pixels by averaging across many exposures as demonstrated in Figure \ref{fig:single_ex}. For both linearity and PTC gain measurements, we apply a linear fit to data from all four light levels.

\begin{figure}[tbp]
\centering
\begin{subfigure}[b]{0.3\textwidth}
\centering
\includegraphics[height=4in]{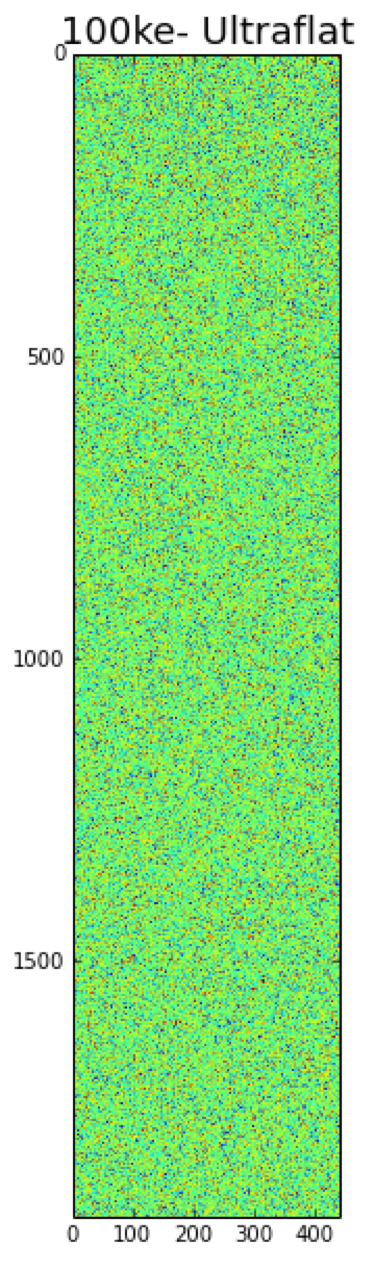}
\caption{Ultraflat image}
\label{fig:p1}
\end{subfigure}
\hfill
\begin{subfigure}[b]{0.3\textwidth}
\centering
\includegraphics[height=4in]{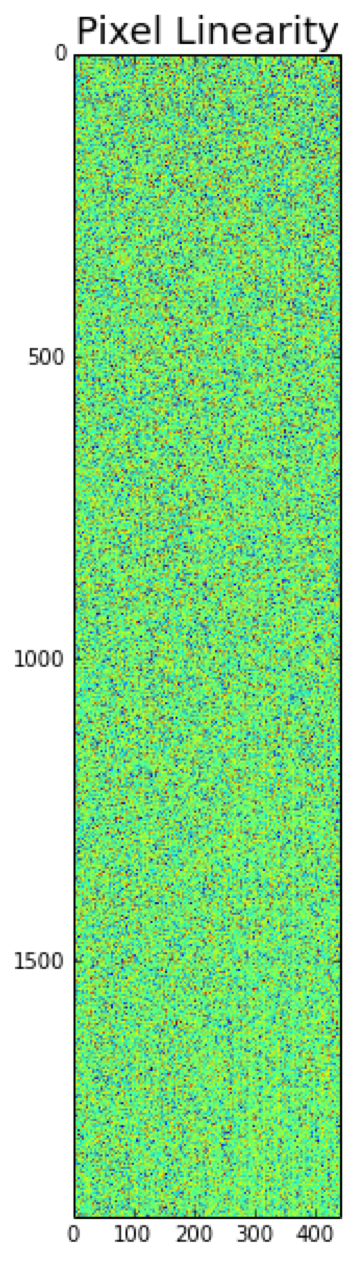}
\caption{Linearity Map}
\label{fig:p2}
\end{subfigure}
\hfill
\begin{subfigure}[b]{0.3\textwidth}
\centering
\includegraphics[height=4in]{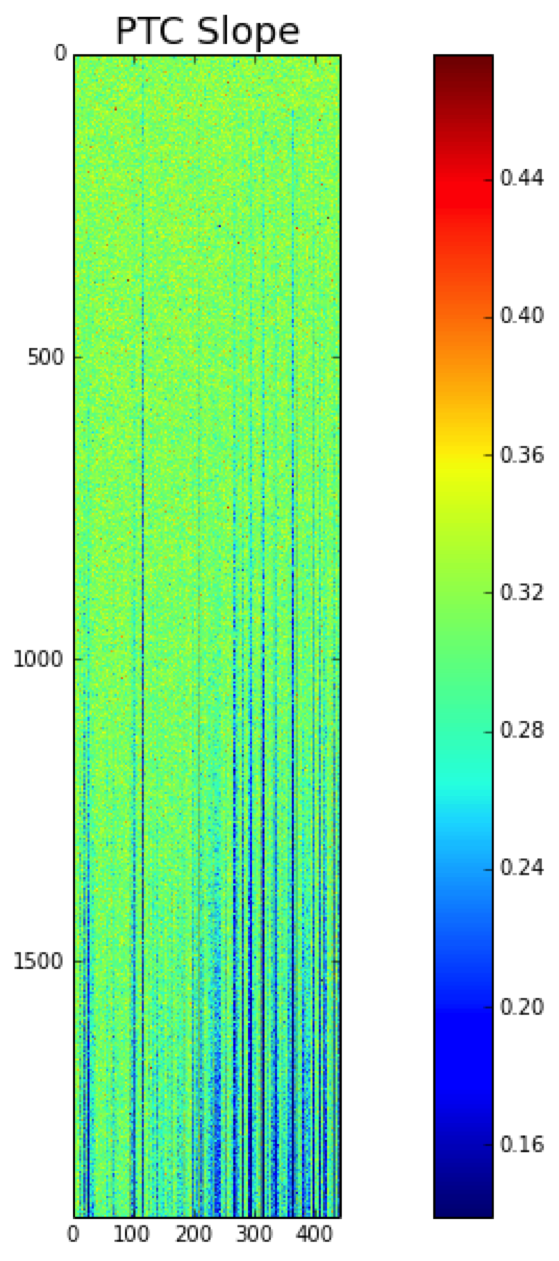}
\caption{PTC Slope Map}
\label{fig:p3}
\end{subfigure}
\caption{These three images summarize the results of our search for spatial structure in single-pixel response properties. The raw 100 ke$^-$ ultraflat shown in (a) has no spatial features, as desired. The map of single-pixel linearities shown in (b) is nearly featureless, with a faint imprint of unexpectedly uniform columns due to the loss of variance observed at the 125 ke$^-$ light level. This columnar loss of variance is much easier to see in (c), where the PTC slopes are highly sensitive to losses of variance.}
\label{fig:maps}
\end{figure}

This single-pixel approach allows us to search for unexpected spatial structure in the linearity and gain response of pixels across the sensor. In Figure \ref{fig:maps}, we see the results of this analysis. In panel \ref{fig:p1}, we see an ultraflat image from one of sixteen amplifier segments, which as discussed previously is featureless. In panel \ref{fig:p2}, we plot a map of the slope of each pixel's mean signal vs. exposure time response. This is again nearly featureless, however, faint columnar structures in the bottom-right corner show that certain columns have an anomalously low shot-to-shot variance. This is seen more clearly in panel \ref{fig:p3}, a map of pixel-by-pixel PTC slope, where the anomalously low variance in particular columns at 125 ke$^-$ pulls the corresponding PTC slopes lower.

It is well-known that bright stars can cause ``blooming'' in CCD sensors when they saturate. However, the data in Figure \ref{fig:p3} are obtained at a flux of 125 ke$^-$, well below the full-well capacity (170 ke$^-$) of the sensor. In addition, blooming would cause a clear signature in the raw image and linearity map, whereas this effect is only clearly observed in variance or gain maps.

\begin{figure}[tbp] 
\centering
\includegraphics[width=4in]{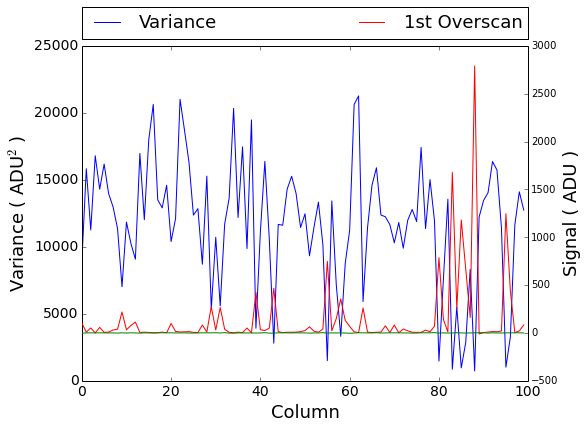}
\caption{A column-by-column empirical correlation between missing column variance and signal measured in each column's nominally-empty first overscan row. Since observing charge in the overscan region is a typical symptom of a CTE problem, we hypothesize that the missing variance effect is due to a flux-dependent CTE effect. There is no charge observed in the other $\sim 10$ rows of the overscan region.}
\label{fig:cte}
\end{figure}

Given that the readout amplifier is in the upper-left-hand corner of each image in Figure \ref{fig:maps}, the loss of variance is clearly proportional to parallel address with respect to this amplifier. Such a dependence leads us to suspect a parallel charge-transfer inefficiency (CTI). In addition, in the columns that exhibit a severe lack of variance, unexpected signal is observed in its corresponding first overscan pixel, as illustrated in Figure \ref{fig:cte}, another typical symptom of CTI. While loss of variance cannot arise in the classic model of charge-transfer inefficiency, a modified charge-transfer efficiency (CTE) model, where the marginal transfer efficiency decreases above some (below full-well) threshold, is a plausible alternative model that could cause loss of variance. However, the value of this threshold, and even the form of this alternative CTE model, are not well-constrained by our data.

Characterizing this atypical flux-dependent CTE problem would require a significant amount of additional data, since the effect depends not only on light level but potentially on other factors such as clock/readout speed. It is unclear why only some columns within an amplifier segment exhibit the effect. Indeed, some amplifiers are entirely free of this effect at the 125ke$^-$ light level. Investigating the turn-on of this effect would require significant amounts of flat fielding data taken at many light levels above 100ke$^-$.

\section{Pixel size variation}

Although the shot noise level in a 100ke$^-$ ultraflat is $0.014\%$, we observe a $0.36\%$ PRNU. Often this would be attributed to variation in quantum efficiency across the silicon and corrected with naive flat-fielding. However, we can use pixel-pixel correlations to assess the impact of both intrinsic pixel size variation and the brighter-fatter effect on PRNU.

\begin{figure}[tbp]
\centering
\begin{subfigure}[b]{0.6\textwidth}
\centering
\includegraphics[width=\textwidth]{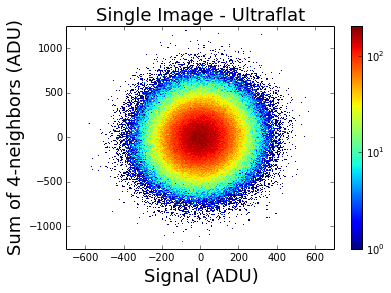}
\caption{In an ultraflat-subracted single image, a modest (4\%) positive correlation between a pixel and the sum of its vertical and horizontal neighbors is observed. This indicates that the observed correlations in flat fields due to the brighter-fatter effect are caused by shot noise as described in the text.}
\label{fig:corr1}
\end{subfigure}
\hfill
\begin{subfigure}[b]{0.6\textwidth}
\centering
\includegraphics[width=\textwidth]{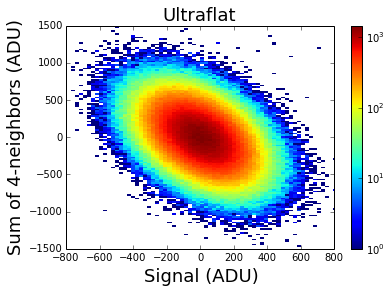}
\caption{The same comparison is shown for an ultraflat image, where a significant (-43\%) anti-correlation is observed, demonstrating the power of ultraflats in highlighting the effects of pixel size variation. }
\label{fig:corr2}
\end{subfigure}
\hfill
\caption{Correlations between pixel flux and sum of 4-neighbor fluxes observed in 100 ke$^-$ data. Each axis has a mean of zero due to the removal of the DC flux term by our high-pass filter.}
\label{fig:corr}
\end{figure}

Subtracting the ultraflat from a single flat-field image highlights the shot noise excursions from each pixel's mean in a single exposure. Due to the excess space charge in the brighter pixels, some converted electrons are redirected toward neighboring pixels. This leads to positive pixel-neighbor correlations in flat fields, as illustrated in In Figure \ref{fig:corr1}. These correlations have been used previously to characterize the brighter-fatter effect in DECam sensors \cite{gruen}.

In Figure \ref{fig:corr2}, the same plot is made for a 100 ke$^-$ ultraflat, where the effects of shot noise and the corresponding brighter-fatter correlations have averaged out, revealing a negative correlation between a pixel's mean flux and its neighbors' mean fluxes. This is indicative of pixel size variation---flux gained by one pixel corresponds to a loss of flux by that pixel's neighbors.

The anisotropy in these ultraflat correlations is measured from correlation coefficients:

\begin{align}
C_{m,n} = \frac{\sum \limits_{i,j} \Phi_{i,j} \Phi_{i+m,j+n}}{\sum \limits_{i,j}\Phi^2_{i,j}}
\end{align}
where $\Phi$ is the flux at a given pixel, the $m,n$ indices represent the displacement from pixel $(i,j)$ to a neighboring pixel of interest, and the $i,j$ indices carry out the sum over all pixels.

\begin{figure}[tbp] 
\centering
\includegraphics[width=4in]{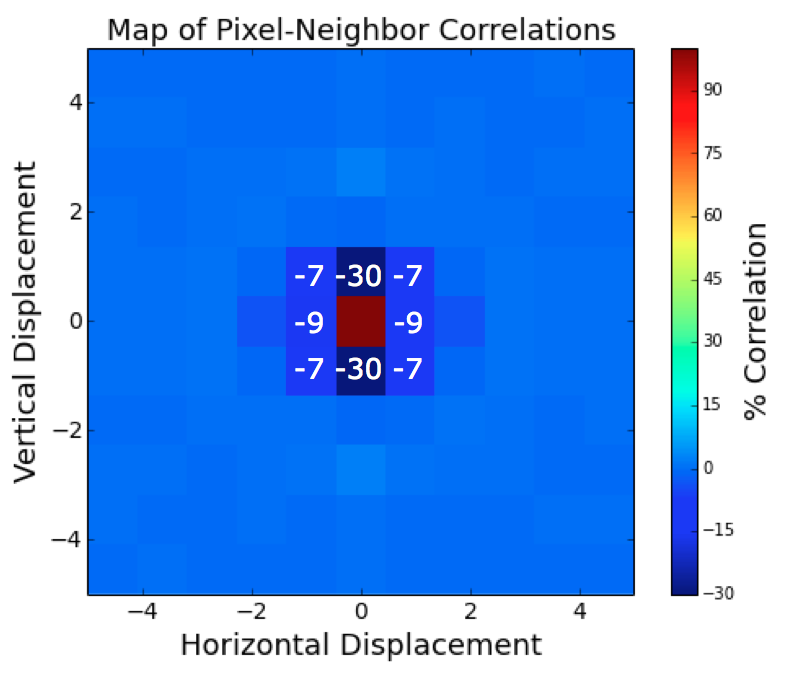}
\caption{A map of pixel-neighbor correlation coefficients computed as described in the text, using data from the 100 ke$^-$ light level.}
\label{fig:corrmap}
\end{figure}

The negative pixel-neighbor correlation coefficients shown in Figure \ref{fig:corrmap} indicate that, in an ultraflat, gains in flux by a particular pixel come at the expense of its neighbors, meaning that ultraflat PRNU can be interpreted as having a significant pixel size variation component. The fact that the anti-correlation is stronger in the column direction is consistent with the hypothesis that these size variations are caused by impurities in the silicon bulk of the sensor---since the clock boundaries are lower potential barriers than the channel stop implants, they should be more easily perturbed by impurities.

The next step in this analysis it to fit an empirical model to this data in order to quantitatively separate the contributions of pixel size variation and local QE variations to PRNU---this will be a subject for further study. However, the clear pixel-neighbor anti-correlations observed in ultraflats set the scale of pixel size variation at the PRNU scale---a few tenths of a percent. Analysis of the potential impact of size variations of this scale on photometry, astrometry, and shape measurements is ongoing.

\section{Conclusions}

Ultraflats are useful probes of pixel response and geometry in CCD sensors. We have used them here to probe the linearity and noise properties of individual pixels, and we confirm the desired absence of spatial structure in pixel response up to a light level of 100 ke$^-$. At 125 ke$^-$, we have described a columnar loss of variance effect we attribute to a flux-dependence of parallel CTE. We have also used pixel-neighbor correlations in ultraflats to observe variations in pixel size in an LSST prototype sensor near the PRNU scale of $\sim .3\%$. Efforts to create a model for these effects (which would allow full decoupling of PRNU contributions from pixel size variation and local QE variation) are ongoing, with the goal of improving the procedure of naïve flat-fielding. We also plan to take images sinusoidal illumination fields, which have the potential to allow improved characterization of structure in pixel size variation across the device.

\acknowledgments

Thanks to Chris Stubbs and Peter Doherty for collecting data and offering helpful advice. This work was performed in part under DOE Contract DE-AC02-76SF00515. This material is based upon work supported by the National Science Foundation Graduate Research Fellowship under Grant No. DGE-114747.


\begin{thebibliography}{9}

\bibitem{science}
Z.~{Ivezic}, et al. {for the LSST Collaboration},
\newblock {\em LSST: from science drivers to reference design and anticipated data
Products}.
\newblock {arXiv:0805.2366}, May 2008.

\bibitem{antilogus}
P.~{Antilogus}, et al.
\newblock {\em The brighter-fatter effect and pixel correlations in CCD sensors}.
\newblock {2014 \em JINST} \textbf{9} C3048.

\bibitem{gruen} 
D.~{Gruen}, et al.
\newblock {\em Characterization and correction of charge-induced pixel shifts in
DECam}.
\newblock {arXiv:1501.02802}, January 2015.

\bibitem{smith}
R.~M. {Smith} and G.~{Rahmer}.
\newblock {\em Pixel area variation in CCDs and implications for precision
photometry}. {\em SPIE
Conf. Ser.} \textbf{7021} (2008) 70212A .

\bibitem{stubbs} 
C.~W. {Stubbs}.
\newblock {\em Precision astronomy with imperfect fully depleted CCDs---an
introduction and a suggested lexicon}.
\newblock {2014 \em JINST} \textbf{9} C3032.

\bibitem{personal}
C. W. Stubbs (personal communication).

\end{thebibliography}
\end{document}